\documentclass[aps, prx, preprint]{revtex4-2}

\usepackage{bm, amsmath, amsfonts, mathtools, physics}
\usepackage[pdftex]{graphicx}
\usepackage{ulem}

\begin{document}

  \title{Weakly nonlinear analysis of Turing pattern dynamics on curved surfaces}

  \author{Ryosuke Nishide${}^{1,}$}
    \email{r-nishide2018@g.ecc.u-tokyo.ac.jp}

  \author{Shuji Ishihara${}^{1,2}$}

  \affiliation{
  ${}^1$~Graduate School of Arts and Sciences, The University of Tokyo, Komaba 3-8-1, Meguro-ku, Tokyo 153-8902, Japan \\
  ${}^2$~Universal Biology Institute, The University of Tokyo, Komaba 3-8-1, Meguro-ku, Tokyo 153-8902, Japan
  }

\begin{abstract}
  Pattern dynamics on curved surfaces are ubiquitous.
  Although the effect of surface topography on pattern dynamics has gained much interest,
  there is a limited understanding of the roles of surface geometry and topology in pattern dynamics.
  Recently,
  we reported that a static pattern on a flat plane can become a propagating pattern on a curved surface [Nishide and Ishihara, Phys. Rev. Lett. 2022].
  By examining reaction-diffusion equations on axisymmetric surfaces,
  certain conditions for the onset of pattern propagation were determined.
  However, this analysis was limited by the assumption that the pattern propagates at a constant speed.
  Here,
  we investigate the pattern propagation driven by surface curvature using weakly nonlinear analysis, which enables a more comprehensive approach to the aforementioned problem.
  The analysis reveals consistent conditions of the pattern propagation similar to our previous results,
  and further predicts that rich dynamics other than pattern propagation, such as periodic and chaotic behaviors, can arise depending on the surface geometry.
  This study provides a new perspective on the relationship between surfaces and pattern dynamics and a basis for controlling pattern dynamics on surfaces.
\end{abstract}

\maketitle

\section{Introduction}
  Pattern formation occurs at a wide range of scales in natural and engineered systems,
  including nanoscale metal stripes, chemical waves, fish and animal skin, vegetation, and atmospheric convection
  ~\cite{murray2003mathmatical, fuseya2021nanoscale, asaba2023growth, manukyan2017living,
   ross1988chemical, tarnita2017theoretical, goyal2021zonal, radja2019pollen}.
  The stages in which pattern formation occurs
  are diverse in terms of spatial dimensions and geometries,
  among which curved surfaces are common and fundamental.
  In particular,
  biological systems are rich sources of pattern dynamics on curved surfaces~\cite{Alonso2016, Bischof2017, tan2020topological, saito2021,
  McGuire2021geographic, maroudas2021topological,wettmann2018an}.
  Recent developments in experimental techniques for measuring and controlling curved surfaces have revealed the functional roles of curved surfaces in pattern formation~\cite{Baptista2019Overlooked, Ehrig2019surface, Schamberger2023curvature}.
  Bin/Amphiphysin/Rvs domain proteins sense the curvature of cellular membranes and regulate the cellular shape~\cite{Antonny2011,simunovic2015when}.
  The distribution of the mechanosensitive protein Piezo1 is regulated by membrane curvature~\cite{yang2022membrane}.
  Cellular migration is guided by the curvature of a substrate
  and is called ``curvotaxis''~\cite{pieuchot2018curvotaxis}.
  The thickness of a cellular sheet depends on the curvature of the substrate~\cite{luciano2021cell}.

  Several theoretical studies have revealed the effects of geometry and topology on pattern formation and dynamics.
  Geodesic curvature causes the splitting~\cite{horibe2019curved},
  rectification~\cite{Davydov2000},
  and swirling~\cite{gomatam1997reaction, yagisita1998spiral, McGuire2021geographic} of excitable waves and variations in the wavefront speed~\cite{bialecki2020traveling, davydov2003critical, kneer2003nucleation}.
  It also controls the position of the localization pattern~\cite{gomez2015phase, singh2022sensing}.
  The rectification of propagating patterns by curved surfaces was also reported in the collective motion of self-propelled particles~\cite{ai2019collective}.
  For pattern dynamics described by polar and nematic variables,
  topological defects are constrained by Poincar\'e--Hopf theorem on closed surfaces~\cite{Brasselet2009Vector},
  and have been investigated in liquid crystals~\cite{Kralj2011,Carenza2022cholesteric},
  flocking~\cite{Shankar2017topological},
  and active nematics~\cite{keber2014topology}.
  This restriction plays a critical role in hydra regeneration,
  where singular points of cortical actin fibers dictate the positional information of morphogenesis~\cite{maroudas2021topological}.

  Turing patterns,
  a prominent example of pattern formation~\cite{turing1952chemical, kondo2010reaction, raspopovic2014digit, ishihara2006turing},
  have also attracted much interest in their dynamics on curved surfaces;
  in fact,
  Turing himself studied the pattern on a sphere to consider embryogenesis~\cite{turing1952chemical}.
  Since then,
  Turing patterns have been studied on several curved surfaces~\cite{Plaza2004, Vandin2016, Charette2019, Krause2021}:
  spheres~\cite{turing1952chemical, varea1999turing, matthews2003pattern, nunez2017diffusion,
  lacitignola2017turing, sanchez2019turing},
  hemispheres~\cite{liaw2001turing, nagata2013reaction},
  tori~\cite{nampoothiri2017role, sanchez2019turing},
  ellipsoids~\cite{nampoothiri2017role, nampoothiri2019effect},
  and deformed axisymmetric cylinders and spheres~\cite{frank2019pinning}.
  These studies have reported that the profile and position of a Turing pattern change to reflect the surface shape.
  For example, Frank et al. studied pattern formation on a cylinder with a ridge,
  and found that the stripe pattern position is modulated by the ridge,
  termed as ``pinning''~\cite{frank2019pinning}.
  All these studies presumed that a Turing pattern,
  which is static on a flat plane, remains static irrespective of the surface geometry.

  We recently reported that a Turing pattern can no longer remain static on curved surfaces~\cite{Nishide2022},
  indicating a new mechanism of pattern propagation caused by surface curvature, i.e.,
  a pattern that is static on a flat plane propagates on a curved surface.
  We primarily studied patterns on axisymmetric surfaces,
  which enabled a tractable analysis.
  Linear stability analysis indicates no sign of Hopf bifurcation,
  similar to the standard Turing instability condition,
  and the growth rate near the uniform state is indistinguishable between flat planes and curved surfaces (see Sec.~\ref{sec:LinearAnalysis}).
  Thus, analysis incorporating nonlinearity is required to understand the onset of such time-dependent dynamics.
  We performed numerical simulations and theoretical analyses based on the relationship between the propagation velocity and the pattern profile (see Eq.~(\ref{eq:velocity_rel}) in Sec.~\ref{sec:propagating_pattern}) and revealed that the onset of propagation is conditioned by symmetries of the surface and pattern.
  The helical pattern propagates unless the surface is reflection-symmetric,
  whereas the parallel pattern is always static (see Fig.~\ref{fig1}).
  These results are generic and were confirmed by numerical simulations of the Brusselator~\cite{Prigogine1977, Pena2001stability}
  and Lengyel--Epstein~\cite{Lengyel1991, Bansagi2015} models.

  In this paper,
  we adopt an alternative analysis method to uncover the manner in which a static pattern on a plane can propagate on a curved surface.
  Our previous analysis assumed that patterns propagate at a constant velocity along the azimuthal direction and used the relationship in Eq.~(\ref{eq:velocity_rel}).
  Instead,
  in this study, we consider the solution of the reaction-diffusion equation (RDE) in the vicinity of the Turing bifurcation point
  and perform a weakly nonlinear analysis.
  Near the bifurcation point,
  the amplitudes of the pattern profile are small,
  which allows us to expand the pattern by the superposition of a few linear modes and derive the amplitude equations~\cite{Cross1993pattern, Pena2001stability, Hoyle2006pattern, kuramoto2012chemical, Charette2019}.
  The obtained amplitude equations indicate that different mode-mode interactions occur
  depending on the symmetries of the surface and pattern profile,
  resulting in either the onset or suppression of pattern propagation.
  This analysis not only leads to consistent results as in \cite{Nishide2022},
  but further predicts that non-trivial dynamics other than propagating patterns,
  including periodic and chaotic behaviors, can appear on curved surfaces.
  These results are confirmed by numerical simulations.

  This paper is organized as follows.
  Sec.~\ref{sec:Model} describes the setup of our problem,
  and summarizes our previous results that the Turing pattern can propagate on a curved surface.
  In Sec.~\ref{sec:AmpEq}, the amplitude equations for the RDE on an axisymmetric surface are discussed.
  Possible forms of amplitude equations depending on the surface and pattern symmetries are determined.
  In Sec.~\ref{sec:AmpEqSte}, the correspondence between the steady-state solutions of the amplitude equations and static and propagating patterns is discussed.
  In Sec.~\ref{sec:AmpEqBr}, the amplitude equations for the Brusselator model are derived and the discussions in the previous sections are numerically confirmed.
  Sec.~\ref{sec:Dynamics} shows that surface curvature can cause diverse pattern dynamics other than propagation,
  which is a summary of the joint
  paper~\cite{joint}.
  Finally, the results are summarized in Sec.~\ref{sec:Discussion}.

\section{Model and Propagating Pattern\label{sec:Model}}

\subsection{Model}
  We study reaction-diffusion systems on an axisymmetric surface represented by $\bm{r}=(x,r(x)\cos\theta,r(x)\sin\theta)$ (Fig.~\ref{fig2}(a)).
  The coordinates $x$ and $\theta$ are defined as $-2\pi \leq x < 2\pi$ and $0\leq \theta < 2\pi$, respectively,
  and periodic boundary conditions are used for both axes unless otherwise mentioned.
  We consider the general form of the RDE as follows:
  \begin{align}
    \partial_t \bm{u} = D\Delta\bm{u} + \bm{R}(\bm{u}), \label{eq:RD}
  \end{align}
  where ${\bm u}$ is the vector representation of the chemical concentrations and a function of the position on the surface $(x,\theta)$ and the time $t$,
  $D$ is a diagonal matrix composed of the diffusion coefficients, $\Delta$ is the Laplace--Beltrami operator,
  and $\bm{R}(\bm{u})$ is the reaction term.
  The Laplace--Beltrami operator $\Delta$ characterizes the effect of the surface curvature on the system~\cite{Krause2019}.
  In general, for a curved surface $\bm{r} = \bm{r}(x^1,x^2)$ parameterized by the coordinates $(x^1, x^2)$, the Laplace--Beltrami operator $\Delta$ for a scalar-field variable is determined by the metric tensor $g_{ij}=\partial_i \bm{r}\cdot\partial_j \bm{r}$,
  where we use the notation $\partial_i \equiv\partial /\partial{x^i}$:
  \begin{align} \label{eq:Laplace-Beltrami_general}
    \Delta\bullet =
        \frac{1}{\sqrt{g}} \partial_{i}\bigg( \sqrt{g}g^{ij} \partial_{j}\bullet \bigg)~,
  \end{align}
  where,
  $g^{ij}$ is the inverse of $g_{ij}$, $g \equiv |\det(g_{ij})|$, and we use the Einstein summation convention for repeated indices.
  For an axisymmetric surface,
  $g_{xx}=1+r'^2$, $g_{\theta\theta} = r^2$, $g_{x\theta}=g_{\theta x} = 0$,
  and $\sqrt{g}=r\sqrt{1+r'^2}$,
  where $r' \equiv dr(x)/dx$.
  The Laplace--Beltrami operator is given by:
  \begin{align}
    \label{eq:Laplace-Beltrami}
    \Delta\bullet =
            \frac{1}{r\sqrt{1+r'^2}} \partial_{x}\bigg( \frac{r}{\sqrt{1+r'^2}} \partial_{x}\bullet\bigg)
            + \frac{1}{r^2} \partial^2_{\theta}\bullet~.
  \end{align}
  We assume that Eq.~(\ref{eq:RD}) has a unique uniform steady solution, $\bm{u}_0$, that satisfies $\bm{R}(\bm{u}_0) = 0$, and the steady-state solution becomes unstable because of the Turing instability indicated by the dispersion relation discussed below.

  As a representative example, we use the Brusselator model given by
  \begin{align}\label{eq:Brusselator}
    \partial_t u &= D_u\Delta u + u^2v - bu - u + a~, \nonumber \\
    \partial_t v &= D_v\Delta v - u^2v + bu~,
  \end{align}
  where $u(t,x,\theta)$ and $v(t,x,\theta)$ represent the chemical concentrations at the position on the surface $(x,\theta)$ and time $t$, $D_u$ and $D_v$ are the diffusion coefficients of $u$ and $v$, respectively, and $a$ and $b$ are model parameters.
  The Brusselator model has a uniform steady solution, $\bm{u}_0 = (a,b/a)^T$,
  and, although it can show various spatiotemporal patterns including static stripes, spiral propagation, and oscillatory patterns~\cite{nazimuddin2022oscillatory}, in this study we selected parameter sets such that the system exhibits a Turing pattern on a flat surface.
  For the numerical data presented in this study, we verified that the pattern is static on a flat plane.
  The details of the numerical simulation method are provided in Supplemental Text Sec.~S1~\cite{Supplemental}.

\subsection{Linear stability analysis \label{sec:LinearAnalysis}}
  Turing instability is characterized by the dispersion relation $\mu(\lambda)$, where $\lambda$ is the eigenvalue of the Laplace--Beltrami operator.
  The dispersion relation $\mu(\lambda)$ is obtained by a linear stability analysis at the uniform state ${\bm u} = {\bm u}_0$, as an eigenvalue of the linear operator
  \begin{align}
  	\mathcal{L} = D \Delta + \partial_{\bm u}{\bm R}(\bm u_0)~,
  \end{align}
  where $\partial_{\bm u}{\bm R}(\bm u_0)$ denotes the Jacobian matrix of ${\bm R}(\bm u)$ at ${\bm u}={\bm u_0}$.
  The eigenfunctions of $\mathcal{L}$ are represented by $\bm{U}_{\lambda,k}(x,\theta) = {\bm A}_{\lambda}W_{\lambda,k}(x,\theta)$, with the Laplace--Beltrami eigenfunction $W_{\lambda,k}(x,\theta)$ defined by $\Delta W_{\lambda, k} = -\lambda W_{\lambda, k}$.
  For flat surfaces, $\lambda$ coincides with the square of the wavenumber.
  For an axisymmetric surface,
  $W_{\lambda, k}$ can be factored into $W_{\lambda,k}(x,\theta) = X_{\lambda, k}(x)e^{ik\theta}$, where $k$ is the wavenumber along the $\theta$-direction and is an integer.
  $X_{\lambda, k}(x)$ is a function of $x \in [-2\pi,~2\pi)$ and is the solution to the following equation:
  \begin{align}
    \frac{r}{\sqrt{1+r'^2}}\partial_x\bigg(\frac{r}{\sqrt{1+r'^2}}\partial_xX_{\lambda, k}\bigg)+\left( \lambda r^2 - k^2\right)X_{\lambda, k}=0~,
  \end{align}
  under the periodic boundary condition.
  Note that the eigenvalue $\lambda$ is a nonnegative real number ($\lambda\in\mathbb{R}_0^+$),
  which is a general feature of the Laplace--Beltrami operator.
  In addition, the eigenvalues are discretized for a closed surface with a finite area.
  The eigenvalues are degenerate two-fold for $k \neq 0$, for which $W_{\lambda,k}(x,\theta)$ and $W_{\lambda,-k}(x,\theta)$ are pairwise eigenfunctions that correspond to the anti-phase shift along the $\theta$-axis.
  Vector ${\bm A}_{\lambda}$ is the eigenvector of the matrix $-\lambda D+\partial_{\bm u}{\bm R}(\bm u_0)$, where $\mathcal{L} {\bm U}_{\lambda,k} = \mu(\lambda) {\bm U}_{\lambda,k}$ is satisfied. The dispersion relation $\mu(\lambda)$ represents the growth rate of a mode characterized by $\lambda$. For Turing instability, $\mu(\lambda)$ takes real positive values within a finite range of $\lambda$, and the corresponding eigenvector $\bm{A}_{\lambda}$ is a real-valued vector.
  Note that the dispersion relation $\mu(\lambda)$, and thus the Turing condition, is the same for any surface.

\subsection{Pattern propagation on axisymmetric surface \label{sec:propagating_pattern}}
  By setting the parameters in the Brusselator model to $(a,b,D_u,D_v) = (2.0, 4.5, 0.5, 1.8)$,
  we obtain the dispersion relation shown in Fig.~\ref{fig2}(b).
  In the numerical simulation of the RDE,
  the initial small fluctuation around the uniform solution $\bm{u}_0$ grows and the chemical concentration eventually forms a spatially periodic pattern (i.e., the Turing pattern).
  The pattern is static on a plane~(Fig.~\ref{fig2}(c)),
  whereas it propagates on an axisymmetric surface,
  as shown in Fig.~\ref{fig2}(d).
  Futhermore, the propagation is observed when the Neumann boundary condition is employed (Fig.~S1~\cite{Supplemental}).

  The velocity of the pattern is related to the pattern profile and surface shape.
  Assuming a traveling wave solution $\bm{u}(x,\rho)$ with $\rho=\theta-\omega t$,
  the angular velocity $\omega$ along the $\theta$-axis is obtained by the following equation:
  \begin{align}
  	\label{eq:velocity_rel}
    \omega = -\frac{\langle \partial_\rho\bm{u},\bm{R}\rangle}{\langle \partial_\rho\bm{u},\partial_\rho\bm{u}\rangle}~,
  \end{align}
  where the inner product is defined by
  \begin{align}
    \langle\bm{\Phi}(x^1\!,x^2),\! \bm{\Psi}(x^1\!,x^2)\rangle \equiv\! \int \! dS~\bm{\Phi}^\dagger(x^1\!,x^2) \bm{\Psi}(x^1\!,x^2)~,
  \end{align}
  with the area element $dS=\sqrt{g}dx^1dx^2$, and $\bm{\Phi}^\dagger$ denoting the Hermitian adjoint vector of $\bm{\Phi}$.

  In our previous study~\cite{Nishide2022},
  we used Eq.~(\ref{eq:velocity_rel}) to reveal the relationship between pattern propagation and the symmetry of patterns and surfaces, as summarized in Fig.~\ref{fig1}.
  The analysis therefore largely depended on the assumption that the pattern propagates at constant speed.
  Note also that we primarily studied axisymmetric surfaces with $2\pi$-periodicity along the $x$-axis;
  in this study,
  we will show that the loss of the $2\pi$-periodicity results in the propagation of the parallel pattern.

\section{Amplitude equation and symmetry of system \label{sec:AmpEq}}

\subsection{Mode decomposition of the pattern dynamics}
  We analyze the propagating pattern on an axisymmetric surface by expanding the pattern using eigenfunctions of the linear operator in a uniform state,
  ${\bm U}_{\lambda,k}(x,\theta) = {\bm A}_{\lambda} W_{\lambda,k}(x,\theta) $.
  If the pattern is not significantly far from the bifurcation point of the Turing instability,
  where the uniform solution ${\bm u}_0$ becomes unstable,
  the pattern profile can be approximated by a linear combination of a few eigenfunctions ${\bm U}_{\lambda,k}$.
  We evaluate the expansion coefficients by numerically calculating the inner product between the eigenfunction ${\bm U}_{\lambda,k}(x,\theta)$ and the propagating pattern
  ${\bm U}(t,x,\theta) \equiv {\bm u}(t,x,\theta)-{\bm u}_0$.

  This expansion was applied to the numerically obtained Brusselator patterns (see Figs.~\ref{fig3} and S2--S4~\cite{Supplemental}).
  The helical propagating pattern shown in Figs.~\ref{fig3}(a) and (b) is well approximated by two pairwise eigenfunctions with the same wavenumber, $k=2$.
  Fig.~\ref{fig3}(c) shows the main part of the expansion:
  \begin{align}\label{eq:ast}
  \bm{U}_\ast(t,x,\theta)&= C_1(t)\bm{U}_{\lambda_1,k}(x,\theta)
  +C_2(t)\bm{U}_{\lambda_2,k}(x,\theta) \\
  &~~~+\bar{C}_1(t)\bm{U}_{\lambda_1,-k}(x,\theta)
  +\bar{C}_2(t)\bm{U}_{\lambda_2,-k}(x,\theta) \nonumber~.
  \end{align}
  This closely approximates the original patterns $U = u(t,x,\theta) - a$ and $ V = v(t,x,\theta)-b/a$.
  Here,
  the complex coefficients $C_{1}$ and $C_{2}$ represent the amplitudes and phases of the corresponding modes, which are given by $|C_i|$ and $\Theta_i=\tan^{-1}\big(\Im(C_i)/\Re(C_i)\big)$, respectively.
  $\bar{C}$ indicates the complex conjugate of $C$.
  As $\bm{u}(t,x,\theta)$ is a real function,
  the coefficients of the pairwise eigenfunction are complex conjugates.
  Notably,
  a single pair of eigenfunctions, $W_{\lambda,\pm k}(x,\theta) = X_{\lambda, \pm k}(x)e^{\pm ik\theta}$,
  is insufficient to express the helical pattern; at least two pairs of eigenfunctions are required.
  In the present example, the two eigenvalues are indicated by the colored dots in the dispersion relation shown in Fig.~\ref{fig3}(d), and the corresponding eigenmodes are shown in Fig.~\ref{fig3}(e).
  As the pattern propagates along the $\theta$-axis, the coefficients of these modes, $C_{1}$ and $C_{2}$,
  oscillate periodically with constant amplitudes $|C_1|$ and $|C_2|$, respectively,
  and with the same frequency $k \omega$,
  where $\omega$ is the angular velocity of the pattern propagation (Figs.~\ref{fig3}(f) and (g)).
  Thus,
  Eq.~(\ref{eq:ast}) can be expressed as
  \begin{align}
    \bm{U}_\ast(t,x,\theta)&=2\big(|C_1|\bm{A}_{\lambda_1}X_1\cos(k(\theta-\omega t) + \theta_1)\nonumber \\
               &\quad+|C_2|\bm{A}_{\lambda_2}X_2\cos(k(\theta-\omega t) + \theta_2)\big)~, \label{eq:Propagating}
  \end{align}
  where $\theta_1$ and $\theta_2$ denote, respectively, the phases of $C_1$ and $C_2$ at $t=0$.
  This expression indicates that the pattern propagates with a constant shape and velocity.

  The static patterns are also well approximated by two pairwise eigenfunctions with the same wavenumber $k$,
  where $C_1$ and $C_2$ are constants (Figs.~S2--S4~\cite{Supplemental}).
  Overall, these results indicate that the time evolution of the propagating pattern can be described by the coefficients $C_i(t)$ for a few modes.

\subsection{Amplitude equations for axisymmetric surfaces}
  Based on the observations earlier, we approximate a pattern on a surface using two pairs of eigenfunctions with the same wavenumber $k$ in the $\theta$-direction as
  \begin{align}
	 \bm{U}(t,x,\theta) &= C_{1} \bm{A}_{\lambda_1} W_{\lambda_1,k} + C_{2} \bm{A}_{\lambda_2}  W_{\lambda_2,k} + c.c.~, \label{eq:U_C1}
  \end{align}
  where $c.c.$ indicates the complex conjugate of the former terms,
  and the coefficients $C_{1}(t)$ and $C_{2}(t)$ are functions of time $t$.
  We explore amplitude equations that describe the pattern dynamics on an axisymmetric curved surface.
  \begin{align}
	 \partial_t C_{1}&=H_{1}(C_{1},C_{2})~,  \label{eq:C1}\\
	 \partial_t C_{2}&=H_{2}(C_{1},C_{2})~.  \label{eq:C2}
  \end{align}
  We mainly consider the situation in which two pairs of real eigenfunctions with the same $k$,
  namely $\bm{A}_{\lambda_1} W_{\lambda_1,k}$ and $\bm{A}_{\lambda_2} W_{\lambda_1,k}$,
  are relevant.
  This situation meets the aforementioned helical pattern and other cases (see Supplemental Text Sec.~S4 for cases in which the two modes have different wavenumbers $k_1$ and $k_2$~\cite{Supplemental}).
  In the vicinity of the Turing bifurcation point,
  the amplitudes of these modes, $|C_{1}|$ and $|C_{2}|$,
  are both expected to be small, and the right-hand sides of the amplitude equations,
  $H_{1}(C_{1},C_{2})$ and $H_{2}(C_{1},C_{2})~$,
  can be expanded with respect to $C_{1}$ and $C_{2}$~\cite{Hoyle2006pattern}.
  The symmetries inherent in the system impose strong restrictions on the possible forms of the amplitude equations,
  as investigated in this section.
  The amplitude equations can also be derived from the original RDE using methods such as the reductive perturbation method~\cite{kuramoto2012chemical};
  we will carry out the derivation for the Brusselator model in Sec.~\ref{sec:AmpEqBr}.

\subsubsection{Amplitude equations for general axisymmetric surfaces \label{sec:generalPAmp}}
  For an axisymmetric surface,
  the RDE is invariant about the translation by arbitrary $p_\theta$ along the $\theta$-direction,
  $\mathcal{T}_\theta$, and the reflection about the $\theta$-direction, $\mathcal{M}_\theta$:
  \begin{align}
    \mathcal{T}_\theta &: \theta\mapsto\theta+p_\theta,~^\forall p_\theta\in\mathbb{R}~,\label{T_tr}\\
    \mathcal{M}_\theta &: \theta\mapsto-\theta~.\label{M_tr}
  \end{align}
  The actions of the transformations $\mathcal{T}_\theta$ and $\mathcal{M}_\theta$ on a pattern (see Eq.~(\ref{eq:U_C1})) can be recast because their actions on the amplitudes are as follows:
  \begin{align}
    \mathcal{T}_\theta&: (C_{1},C_{2}) \mapsto (C_{1}e^{-ikp_\theta},C_{2}e^{-ikp_\theta})~,\\
    \mathcal{M}_\theta&: (C_{1},C_{2}) \mapsto (\bar{C}_{1},\bar{C}_{2})~.
  \end{align}
  The amplitude equations, which are the reduction equations of RDE,
  should also be invariant to these transformations.
  Thus, the general form of the amplitude equations up to the third-order expansion of $C_{i}$ is given by
  \begin{align}
    \partial_tC_{1}&=a_1C_{1}+a_2|C_{1}|^2C_{1}+a_3|C_{2}|^2C_{2}+a_4|C_{1}|^2C_{2} \nonumber \\
    &\quad+a_5|C_{2}|^2C_{1}+a_6C_{1}^2\bar{C}_{2}+a_7C_{2}^2\bar{C}_{1}~, \label{eq:amp_C1}\\
    \partial_tC_{2}&=b_1C_{2}+b_2|C_{1}|^2C_{1}+b_3|C_{2}|^2C_{2}+b_4|C_{1}|^2C_{2} \nonumber \\
    &\quad+b_5|C_{2}|^2C_{1}+b_6C_{1}^2\bar{C}_{2}+b_7C_{2}^2\bar{C}_{1} ~,\label{eq:amp_C2}
  \end{align}
   where the coefficients $a_j$ and $b_j$ are real values owing to the invariance of the transformation $\mathcal{M}_\theta$.
   The second-order terms do not appear in the equations.
   The coefficients $a_1$ and $b_1$ coincide with growth rates $\mu(\lambda_1)$ and $\mu(\lambda_2)$, respectively.
   The other coefficients are not determined solely by the discussion of symmetries given here.
   Below, we assume that the coefficients are chosen such that the solution does not diverge;
   otherwise,
   higher-order terms are required to prevent divergence.

   Using the complex variables $C_{1}$ and $C_{2}$ as $C_{1}=\eta e^{i\phi},~C_{2}=\xi e^{i\psi}$,
   the pattern profile in Eq.~(\ref{eq:U_C1}) can be written as
   \begin{align}
      \bm{U}(t,x,\theta)&=2\eta\bm{A}_{\lambda_1}X_{\lambda_1,k}\cos(k\theta+\phi)\nonumber \\
      &\quad+2\xi\bm{A}_{\lambda_2}X_{\lambda_2,k}\cos(k\theta+\psi)~,
   \end{align}
   where the amplitudes, $\eta$ and $\xi$, and phase variables, $\phi$ and $\psi$, are real-valued functions of time $t$.
   The amplitude equations can be rewritten as follows:
   \begin{align}
    \partial_t\eta &= \big(a_1+a_2\eta^2+a_5\xi^2\big)\eta + a_7\eta\xi^2\cos(2\alpha) \nonumber \\
    &\quad+\big(a_3\xi^2+(a_4+a_6)\eta^2\big)\xi\cos(\alpha)~, \label{eq:eta_general} \\
    \partial_t\xi &= \big(b_1+b_3\xi^2+b_4\eta^2\big)\xi + b_6\eta^2\xi\cos(2\alpha) \nonumber \\
    &\quad+\big(b_2\eta^2+(b_5+b_7)\xi^2\big)\eta\cos(\alpha)~, \label{eq:xi_general}\\
    \eta\xi \partial_t\alpha &= -\big(a_3\xi^4+(a_4-a_6+b_5-b_7)\eta^2\xi^2+b_2\eta^4 \nonumber\\
    &\qquad\quad+2(a_7\xi^2+b_6\eta^2)\eta\xi\cos(\alpha)\big)\sin(\alpha)~, \label{eq:alpha_general}
   \end{align}
   where we introduce the phase difference $\alpha \equiv \phi-\psi$.
   These equations are closed in three variables, $\eta$, $\xi$, and $\alpha$,
   because the system is invariant for translation along the $\theta$-axis.
   Note also that they are invariant under the transformation $\alpha \rightarrow -\alpha$ because the system has reflection-symmetry along the $\theta$-axis.
   The phase variables $\phi$ and $\psi$ evolve according to the following equations:
   \begin{align}
    \eta\partial_t\phi&=-\xi((a_4-a_6)\eta^2+a_3\xi^2)\sin(\alpha) \nonumber \\
                      &\quad-a_7\eta\xi^2\sin(2\alpha),\label{eq:phi_general}\\
    \xi\partial_t\psi&=+\eta(b_2\eta^2+(b_5-b_7)\xi^2)\sin(\alpha) \nonumber \\
                      &\quad+b_6\eta^2\xi\sin(2\alpha).\label{eq:psi_general}
   \end{align}
   Eqs.~(\ref{eq:eta_general})--(\ref{eq:alpha_general}),
   supplemented by Eqs.~(\ref{eq:phi_general}) and (\ref{eq:psi_general}), constitute the general form of the amplitude equations for pattern dynamics on an axisymmetric surface.
   Note that the derived amplitude equations Eqs.~(\ref{eq:amp_C1}) and (\ref{eq:amp_C2})
   are also valid for non-Turing systems
   as long as the system satisfies similar conditions such as invariance to the transformations Eqs.~(\ref{T_tr}) and (\ref{M_tr}).

\subsubsection{Amplitude equations for reflection-symmetric surfaces \label{sec:reflectionsymmetricAmplitude}}
  An axisymmetric surface can have additional symmetries.
  Here, we consider the case in which the surface is reflection-symmetric about the $x$-axis,
  satisfying $r(x) = r(-x)$,
  where the origin of the $x$-axis can be arbitrary because of the periodic boundary condition.
  For such a surface, RDE is invariant under the reflection about the $x$-direction, $\mathcal{M}_x$:
  \begin{align}
    \mathcal{M}_x&: x\mapsto-x~.
  \end{align}
  The eigenfunctions of the Laplace--Beltrami operator $X_{\lambda,k}(x)$ are either even or odd with respect to $x$.
  The action of the reflection $\mathcal{M}_x$ on $C_{i}$ is dependent on the parity of $X_{\lambda_i,k}(x)$ and acts as
  \begin{align}
     \mathcal{M}_x&: C_{i} \mapsto \left\{
     \begin{array}{cl}
     +C_{i} &~\mbox{if $X_{\lambda_i,k}$ is an even function}~,\\
     -C_{i} &~\mbox{if $X_{\lambda_i,k}$ is an odd function}~.
     \end{array}
     \right.
  \end{align}
  When $X_{\lambda_1,k}(x)$ and $X_{\lambda_2,k}(x)$ are both even or odd functions, the amplitude equations are the same as those in the general case of Eqs.~(\ref{eq:amp_C1}) and (\ref{eq:amp_C2}).
  However, when $X_{\lambda_1,k}(x)$ and $X_{\lambda_2,k}(x)$ are a combination of even and odd functions, the amplitude equations are reduced to
  \begin{align}
    \partial_tC_{1}&=a_1C_{1}+a_2|C_{1}|^2C_{1}+a_5|C_{2}|^2C_{1}+a_7C_{2}^2\bar{C}_{1}~, \label{eq:C1_RS} \\
    \partial_tC_{2}&=b_1C_{2}+b_3|C_{2}|^2C_{2}+b_4|C_{1}|^2C_{2}+b_6C_{1}^2\bar{C}_{2}~, \label{eq:C2_RS}
  \end{align}
  corresponding to the case $a_3=a_4=a_6=0$ and $b_2=b_5=b_7=0$ in Eqs.~(\ref{eq:amp_C1}) and (\ref{eq:amp_C2}). The amplitude and phase variables obey
  \begin{align}
    \partial_t\eta&=\big(a_1+a_2\eta^2+a_5\xi^2\big)\eta+a_7\eta\xi^2\cos(2\alpha)~,  \label{eq:eta_reflection}\\
    \partial_t\xi&=\big(b_1+b_3\xi^2+b_4\eta^2\big)\xi+b_6\eta^2\xi\cos(2\alpha)~,\\
    \partial_t\alpha&=-(a_7\xi^2+b_6\eta^2)\sin(2\alpha) \label{eq:alpha_reflection}~,
  \end{align}
  and
  \begin{align}
    \partial_t\phi&=-a_7\xi^2\sin(2\alpha)~, \label{eq:phi_reflection}\\
    \partial_t\psi&=+b_6\eta^2\sin(2\alpha)~ \label{eq:psi_reflection}.
  \end{align}

  \subsubsection{Amplitude equations for $2\pi$-periodic surface along the $x$-axis \label{sec:2piPAmp}}
  The axisymmetric surface can be periodic along the $x$-axis, satisfying $r(x) = r(x+p_x)$,
  where $p_x$ is a constant smaller than the axial length of the surface.
  Here,
  we consider a simple case in which the surface is $2\pi$-periodic  (as in our previous study~\cite{Nishide2022}),
  satisfying $r(x) = r(x+2\pi)$,
  and the domain of $x$ is defined as $-2\pi \leq x < 2\pi$.
  For such a surface,
  the RDE is invariant to the $2\pi$-translation along the $x$-axis and $\mathcal{T}_{x,2\pi}$ is defined as
  \begin{align}
    \mathcal{T}_{x,2\pi} &: x\mapsto x+2\pi~,
  \end{align}
  and reflection about the $\theta$-direction, $\mathcal{M}_{\theta}$.
  According to Bloch's theorem~\cite{kittel2018introduction},
  the eigenfunctions of the Laplace--Beltrami operator $X_{\lambda,k}(x)$ either change sign or remain unchanged by the action of $\mathcal{T}_{x,2\pi}$.
  \begin{align}
    X_{\lambda,k}(x+2\pi) = \pm X_{\lambda,k}(x).
  \end{align}
  The action of $2\pi$-translation $\mathcal{T}_{x,2\pi}$ on $C_{i}$ act as
  \begin{align}
     \mathcal{T}_{x,2\pi}&: C_{i} \mapsto \left\{
     \begin{array}{cl}
     +C_{i} &~\mbox{if $X_{\lambda_i,k}(x+2\pi)=+X_{\lambda_i,k}(x)$},\\
     -C_{i} &~\mbox{if $X_{\lambda_i,k}(x+2\pi)=-X_{\lambda_i,k}(x)$}
     \end{array}
     \right.
  \end{align}
  When $X_{\lambda_1,k}(x)$ and $X_{\lambda_2,k}(x)$ both undergo the same transformation by $\mathcal{T}_{x,2\pi}$,
  the amplitude equations are the same as those in the general case of Eqs.~(\ref{eq:amp_C1}) and (\ref{eq:amp_C2}).
  Conversely,
  when $X_{\lambda_1,k}(x)$ and $X_{\lambda_2,k}(x)$ undergo transformations of different signs by $\mathcal{T}_{x,2\pi}$,
  the amplitude equations are the same as those on the reflection-symmetric surface in Eqs.~(\ref{eq:C1_RS}) and (\ref{eq:C2_RS}).

\subsubsection{Patterns consisting of single paired modes \label{sec:single_mode}}
  The patterns can be dominated by only a single paired mode,
  as exemplified in Fig.~\ref{fig4},
  where the amplitudes of the other modes decrease.
  \begin{align}
    \bm{U} = C\bm{A}_{\lambda}X_{\lambda,k}e^{ik\theta}+c.c.~.
  \end{align}
  The pattern is always reflection-symmetric around the $\theta$-axis as $U(x,\theta)=U(x,-\theta)$
  where the origin of the $\theta$-axis is properly chosen.
  Based on the invariance by the transformations $\mathcal{T}_\theta$ and $\mathcal{M}_\theta$,
  the amplitude equation is given by
  \begin{align}
    \partial_tC &= a_1C+a_2|C|^2C ~,
  \end{align}
  where $a_1$ coincides with the growth rate $\mu(\lambda)$ and $a_2$ is supposed to be a negative real value.
  With $C = \eta e^{i \phi}$,
  we obtain:
  \begin{align}
    \partial_t\eta &= a_1\eta+a_2\eta^3~,\\
    \partial_t\phi &= 0~. \label{eq:signle_phase_vel}
  \end{align}
  These equations indicate that the pattern does not exhibit propagation.

\subsubsection{Patterns on a flat surface \label{sec:flatsurface}}
  Summarizing the analysis for the flat surface case is helpful.
  On the $xy$-plane, the pattern can be expanded by using the Fourier mode $e^{i(k_xx+k_yy)}$.
  The stripe pattern resulting from Turing instability (Fig.~\ref{fig2}) comprises at most %composed of
  two paired modes and is approximated by
  \begin{align}
    \bm{U} = (C_1e^{i(k_xx+k_yy)}+C_2e^{i(k_xx-k_yy)})\bm{A}+c.c.~.
  \end{align}
  The amplitude equations are derived by considering that the
  plane is invariant to translations and reflections:
  \begin{align}
    \partial_tC_1 &= a_1 C_1 + a_2|C_1|^2C_1 + a_5|C_2|^2C_1~, \label{eq:AmpFlat1}\\
    \partial_tC_2 &= b_1 C_2 + b_3|C_2|^2C_2 + b_4|C_1|^2C_2~. \label{eq:AmpFlat2}
  \end{align}
  Then, amplitudes and phases obey
  \begin{align}
    \partial_t\eta&=a_1\eta+a_2\eta^3+a_5\eta\xi^2~,\\
    \partial_t\xi&=b_1\xi+b_2\xi^3+b_4\eta^2\xi~,\\
    \partial_t\phi&=\partial_t\psi=0~,  \label{eq:phase_flatsurface}
  \end{align}
  and pattern propagation is not possible.
  In the case of a plane, coupling with higher-order harmonic modes
  such as $e^{ik_xx}$ and $e^{i2k_xx}$ can result in propagating waves,
  even in a one-dimensional (1D) system~\cite{Hoyle2006pattern}.
  We ignore such situations because they generally occur far from the Turing bifurcation point and depend on the specific nature of the system,
  such as domain size (see Supplemental Text Sec.~S4~\cite{Supplemental}).

\section{Steady-state solutions of amplitude equations \label{sec:AmpEqSte}}
   Here, we investigate the steady-state solutions of the amplitude equations derived in the preceding section: $(\eta,\xi, \alpha) =(\eta_0, \xi_0, \alpha_0)$.
   For the given solution,
   the symmetries of the corresponding patterns and phase velocities $\partial_t\phi, \partial_t \psi$ are determined.
   As a pattern of single paired modes cannot propagate (see Eq.~(\ref{eq:signle_phase_vel})),
   we consider the case in which two pairs of modes are relevant and $\eta_0$ and $\xi_0$ are assumed to be non-zero.

\subsection{Steady-state solutions of Eqs.~(\ref{eq:eta_reflection})--(\ref{eq:alpha_reflection})}
  The amplitude equations (\ref{eq:eta_reflection})--(\ref{eq:alpha_reflection}) are obtained for the reflection-symmetric surface with eigenfunctions that have different parities as $X_1(-x)=X_1(x)$ and $X_2(-x)=-X_2(x)$ (Sec.~\ref{sec:reflectionsymmetricAmplitude}),
  or for a $2\pi$-periodic surface with eigenfunctions that transform differently as $X_1(x+2\pi)=X_1(x)$ and $X_2(x+2\pi)=-X_2(x)$ (Sec.~\ref{sec:2piPAmp}).
  In the steady-state,
  either of the following relationships should hold from Eq.~(\ref{eq:alpha_reflection}):
  \begin{align}
       &\alpha_0=\frac{n\pi}{2}~,~n\in\mathbb{Z}~, \label{sol:reflec1} \\
       &a_7\xi_0^2+b_6\eta_0^2=0~. \label{sol:reflec2}
  \end{align}

  For the solution $\alpha_0 = n\pi/2$ (Eq.~(\ref{sol:reflec1})),
  the phase equations in Eqs.~(\ref{eq:phi_reflection}) and (\ref{eq:psi_reflection}) satisfy
  $\partial_t\phi=\partial_t\psi=0$, indicating that this solution represents a static pattern.
  The profile of the pattern is expressed as
  \begin{widetext}
  \begin{align}
    \label{eq:SteadyU}
    {\bm U}(x,\hat{\theta}) &=
    2\eta_0\bm{A}_{\lambda_1}X_{\lambda_1,k}(x)\cos \bigl(\hat{\theta}\bigr)
    +2\xi_0\bm{A}_{\lambda_2}X_{\lambda_2,k}(x)\cos \bigl(\hat{\theta}-\alpha_0 \bigr)~\\
    &=
    \begin{dcases}
      2\eta_0\bm{A}_{\lambda_1}X_{\lambda_1,k}(x)\cos\bigl(\hat{\theta}\bigr)
      +2(-1)^{n-1}\xi_0\bm{A}_{\lambda_2}X_{\lambda_2,k}(x)\cos\bigl(\hat{\theta}\bigr)
      & \text{if $\alpha_0 = n \pi$}  \\
     2\eta_0\bm{A}_{\lambda_1}X_{\lambda_1,k}(x)\cos \bigl(\hat{\theta}\bigr)+2(-1)^{n-1}\xi_0\bm{A}_{\lambda_2}X_{\lambda_2,k}(x)\sin \bigl(\hat{\theta}\bigr)
     & \text{if $\alpha_0 = (n + 1/2)\pi$}
    \end{dcases}
    \label{eq:SteadyU2}
 \end{align}
 \end{widetext}
  where $\hat{\theta}=k\theta+\phi_0$ and $\phi_0$ is a constant determined by the initial condition.
  When $\alpha_0 = n \pi$, the pattern is reflection-symmetric along the $\theta$-axis,
  satisfying ${\bm U}(x,\hat{\theta}) = {\bm U}(x,-\hat{\theta})$.
  Fig.~\ref{fig5}(a) shows an example of a $2\pi$-periodic surface,
  where we set $r(x)=2+0.3\cos(x)+0.2\sin(x)$.
  In the other case, $\alpha_0 = (n + 1/2)\pi$, the pattern symmetry depends on the surface symmetry.
  On reflection-symmetric surfaces,
  the pattern is inversion-symmetric, satisfying ${\bm U}(x,\hat{\theta}) = {\bm U}(-x,-\hat{\theta})$.
  On $2\pi$-periodic surfaces,
  the pattern is invariant against the simultaneous transformation of reflection about the $\theta$-axis and $2\pi$-translation along the $x$-axis,
  satisfying ${\bm U}(x,\hat{\theta}) = {\bm U}(x+2\pi,-\hat{\theta})$.
  Fig.~\ref{fig5}(b) shows an example of the case on a reflection-symmetric surface ($r(x)=2.2+0.2\cos(x)+0.4\cos(x/2)$).

  For the steady-state solution given by Eq.~(\ref{sol:reflec2}),
  the phase velocities $\partial_t \phi$ and $\partial_t \psi$ are generally non-zero and constant ($\partial_t \phi = \partial_t \psi \neq 0$),
  indicating a propagating pattern.
  The pattern shows neither reflection- nor inversion- symmetry because of the superposition of the two paired modes in $\alpha_0\neq n\pi/2$.
  This non-trivial solution was omitted in the previous approach~\cite{Nishide2022}.
  Fig.~\ref{fig5}(c) shows an example of a propagating pattern on a $2\pi$-periodic surface
  ($r(x)=1.9+0.21\cos(x)+0.06\sin(2x)$).
  This non-trivial solution is absent on a flat-plane where $b_6 = a_7=0$ (see Eqs.~(\ref{eq:AmpFlat1}) and (\ref{eq:AmpFlat2})).

\subsection{Steady-state solutions of Eqs.~(\ref{eq:eta_general})--(\ref{eq:alpha_general})}
  Next, we consider the steady-state solutions of the general amplitude equations given by Eqs.~(\ref{eq:eta_general})--(\ref{eq:alpha_general}).
  The equations are obtained for a general axisymmetric surface (Sec.~\ref{sec:generalPAmp}),
  for cases in which
  $X_1(x)$ and $X_2(x)$ are both even or odd functions on a reflection-symmetric surface (Sec.~\ref{sec:reflectionsymmetricAmplitude}),
  or when $X_1(x)$ and $X_2(x)$ are transformed with the same sign by $2\pi$-translation along the $x$-axis on a $2\pi$-periodic surface (Sec.~\ref{sec:2piPAmp}).
  In the steady-state,
  Eq.~(\ref{eq:alpha_general}) yields to either of the following solutions:
  \begin{align}
    &\alpha_0=n\pi,~n\in\mathbb{Z}, \label{sol:general1}\\
    &\cos \alpha_0=
    -\frac{a_3\xi_0^4+(a_4-a_6+b_5-b_7)\eta_0^2\xi_0^2+b_2\eta_0^4}{2(a_7\xi_0^2+b_6\eta_0^2)\eta_0\xi_0}~, \label{sol:general2} \\
    & a_3\xi_0^4+(a_4-a_6+b_5-b_7)\eta_0^2\xi_0^2+b_2\eta_0^4 = 0  \nonumber\\
    &~~~~~~~~~~~~~~~~~~~~~~~~~~~~~~ \mbox{with}~~ a_7\xi_0^2+b_6\eta_0^2=0~. \label{sol:general3}
  \end{align}
  From Eqs.~(\ref{eq:phi_general}) and (\ref{eq:psi_general}),
  the first solution $\alpha_0=n\pi$ represents a static pattern
  whereas the others represent propagating patterns.
  Note that at $a_3 = a_4 = a_6 = b_2 = b_5 = b_7 = 0$, the solutions Eqs.~(\ref{sol:general2}) and (\ref{sol:general3}) coincide with the solutions $\alpha_0=(n+1/2)\pi$ included in Eq.~(\ref{sol:reflec1}) and Eq.~(\ref{sol:reflec2}), respectively.

  For the solution $\alpha_0=n\pi$,
  the static pattern takes the form of the first line of Eq.~(\ref{eq:SteadyU2}) and is reflection-symmetric about the $\theta$-axis.
  The pattern has an additional symmetry depending on the surface symmetry and combination of eigenfunctions.
  When both $X_1(x)$ and $X_2(x)$ are even functions on the reflection-symmetric surface,
  the pattern has symmetry ${\bm U}(x,\hat{\theta}) = {\bm U}(x,-\hat{\theta}) = {\bm U}(-x,\hat{\theta}) = {\bm U}(-x,-\hat{\theta})$, while when both are odd,
  it has the symmetry ${\bm U}(x,\hat{\theta}) = {\bm U}(x,-\hat{\theta}) = {\bm U}(-x,\hat{\theta}+\pi) = {\bm U}(-x,-\hat{\theta}+\pi)$.
  An example of the latter static pattern is shown in Fig.~\ref{fig6}(a) ($r(x)=1.95+0.21\cos(x)-0.06\cos(2x)+0.015\cos(x/2)$).
  On a $2\pi$-periodic surface,
  when $X_1(x+2\pi) = X_1(x)$ and $X_2(x+2\pi) = X_2(x+2\pi)$,
  the pattern has the symmetry ${\bm U}(x,\hat{\theta}) = {\bm U}(x,-\hat{\theta}) = {\bm U}(x+2\pi,\hat{\theta}) = {\bm U}(x+2\pi,-\hat{\theta})$,
  while when $X_1(x+2\pi) = -X_1(x)$ and $X_2(x+2\pi) = -X_2(x+2\pi)$,
  it has ${\bm U}(x,\hat{\theta}) = {\bm U}(x,-\hat{\theta}) = {\bm U}(x+2\pi,\hat{\theta}+\pi) = {\bm U}(x+2\pi,-\hat{\theta}+\pi)$.
  An example of the latter static pattern is shown in Fig.~\ref{fig6}(b) ($r(x)=1.8+0.15\cos(x)+0.09\cos(2x)$).

  For other solutions, Eqs.~(\ref{sol:general2}) and (\ref{sol:general3}),
  $\alpha_0$ takes a non-trivial value, and the corresponding propagating patterns have no symmetry. Fig.~\ref{fig6}(c) presents an example for the solution of Eq.~(\ref{sol:general2}) on a general axisymmetric surface ($r(x)=2.2+0.2\cos(x)+0.4\cos(x/2-\pi/4)$).
  Regarding the solution Eq.~(\ref{sol:general3}),
  the parameters are required to
  satisfy additional conditions, such as $a_3b_6^2 + b_2a_7^2 - a_7b_6 (a_4 - a_6 + b_5 - b_7) = 0$,
  which are unlikely to be achieved.
  %\magenta{\sout{and are unlikely to be realized.}}

\subsection{Classification of parallel and helical patterns by steady-state solution of amplitude equations}
  The dynamics of the parallel and helical patterns (Fig.~\ref{fig1}) can be understood using the steady-state solutions discussed above.
  On a $2\pi$-periodic surface along the $x$-axis,
  parallel patterns are static,
  whereas helical patterns are static only on a reflection-symmetric surface (Fig.~\ref{fig7} (a)--(c)).
  These patterns correspond to the steady-state solution $\alpha_0 = \pi/2$ in Eq.~(\ref{sol:reflec1}),
  %\red{with each composed}
  each consisting of eigenfunctions with characteristic symmetry;
  for instance,
  the helical pattern on a reflection-symmetric surface is composed of eigenfunctions with $X_1(x)=X_1(-x)=X_1(x+2\pi)$ and $X_2(x)=-X_2(-x)=X_2(x+2\pi)$.
  The helical pattern begins to move because of
  the loss of reflection-symmetry \cite{Nishide2022},
  which corresponds to the steady-state solution in Eq.~(\ref{sol:general2}) of the amplitude equations in Eqs.~(\ref{eq:eta_general})--(\ref{eq:alpha_general}),
  where $X_1(x)=X_1(x+2\pi)$, $X_2(x)=-X_2(x+2\pi)$ and the phase difference $\alpha_0 \neq n\pi$ (Fig.~\ref{fig7}(d)).

  The parallel static pattern on a $2\pi$-periodic surface along the $x$-axis satisfies $X_1(x)=X_1(x+2\pi)$ and $X_2(x)=-X_2(x+2\pi)$.
  Because of the loss of the $2\pi$-periodicity of the surface,
  these relationships no longer hold,
  and the solution of the amplitude equations becomes Eq.~(\ref{sol:reflec1}),
  indicating the onset of a propagating wave in a parallel pattern.
  A numerical example is shown in Fig.~\ref{fig8}.

\subsection{Summary of amplitude equations and steady-state solutions}
  We derived the amplitude equations~(\ref{eq:amp_C1}) and (\ref{eq:amp_C2})
  for a general axisymmetric surface on which two pairs of modes couple with each other,
  and the static and propagating patterns correspond to the steady solutions of the amplitude equations.
  Some coupling terms are eliminated because of the symmetry of the surface (see Eqs.~(\ref{eq:C1_RS}) and (\ref{eq:C2_RS})),
  which suppresses pattern propagation.
  The propagation of helical and parallel patterns is caused by the loss of reflection-symmetry and $2\pi$-periodicity, respectively, of the surface along the $x$-axis.
  For a flat surface,
  the amplitude equations are reduced to Eqs.~(\ref{eq:AmpFlat1}) and (\ref{eq:AmpFlat2}),
  and the pattern is always static.
  The relationships obtained for the pattern dynamics and the symmetry of the surface and pattern are consistent with those reported in a previous study~\cite{Nishide2022}.
  In addition,
  from the amplitude equations, we obtained a non-trivial solution (Eq.~(\ref{sol:reflec2})) for which the pattern generally has no specific symmetry and propagates.
  This solution allows the Turing pattern to propagate even on highly symmetric surfaces.
  A propagating pattern on a sphere was observed in numerical simulation~\cite{Nishide2022}.

\section{Amplitude equation for the Brusselator model \label{sec:AmpEqBr}}

\subsection{Reductive perturbation method}
  The amplitude equations for  RDE on axisymmetric curved surfaces can be derived by applying the reductive perturbation method to reaction-diffusion systems in the vicinity of the Turing bifurcation. Consider the general RDE expressed in Eq.~(\ref{eq:RD}),
  for which we explicitly include a bifurcation parameter $\nu$ in the reaction term as $\bm{R}({\bm u}, \nu)$.
  As before, we assume that the equation has a unique uniform steady solution, $\bm{u}_0$, satisfying $\bm{R}(\bm{u}_0, \nu) = 0$.
  The uniform solution becomes unstable for $\nu>0$, and a Turing pattern occurs.

  In the vicinity of the Turing bifurcation at a positive, sufficiently small value of $\nu$,
  the amplitude of the pattern, $|{\bm U}| \equiv |{\bm u}-{\bm u}_0|$,
  is expected to be small, and expanding the RDE by $\bm{U}$ and $\nu$ is feasible.
  The amplitude equations can then be derived from the RDE;
  see the details of the derivation in Supplemental Text Sec.~S2~\cite{Supplemental}.
  In summary, the pattern dynamics of the RDE are approximated by a lower-order expansion with respect to $\nu$ and amplitude $C_i$,
  and the pattern dynamics are dominated by the time evolution of $C_i$:
  \begin{align}
    \bm{U}(t)&=\bm{U}_0(\{C_i(t)\})+\bm{\rho}(\{C_i(t)\})~,\\
    \partial_t C_i(t)&=H_i(\{C_i(t)\})~,
  \end{align}
  where $\bm{U}_0$ and $\bm{\rho}$ are the first- and higher-order terms in the amplitudes, respectively,
  and the second equation is referred to as the amplitude equation.
  These are derived in Eq.~(S41) and in Eq.~(S42), respectively.
  The obtained equation (S42) not only validates Eqs.~(\ref{eq:amp_C1}) and (\ref{eq:amp_C2}),
  which are derived as a general form of the amplitude equations,
  but also determines their coefficients.
  The simpler forms of amplitude equations~(\ref{eq:C1_RS}) and (\ref{eq:C2_RS}) can be understood as follows.
  As shown in Eqs.~(S44)--(S49),
  the coefficients include integrals of the products of the Laplace--Beltrami eigenfunctions $X_{\lambda_i,k}$.
  For example, if the surface is reflection-symmetric about the $x$-axis and $X_{1}(x)$ and $X_{2}(x)$ are a pair of even and odd functions,
  some coefficients are equal to zero owing to the combination of the parity of the eigenfunctions in the integrands, and Eq.~(S42) coincides with Eqs.~(\ref{eq:C1_RS}) and (\ref{eq:C2_RS}).

\subsection{Derivation of Amplitude equation for Brusselator model \label{sec:Brusselator_Reduction}}

  We derive the amplitude equations for the Brusselator model in Eqs.~(\ref{eq:amp_C1}) and (\ref{eq:amp_C2}) following the general procedure described in Supplemental Text Sec.~S2~\cite{Supplemental}.
  We choose the bifurcation parameter $\nu = b-b_0$,
  where the reference parameters of the reductive perturbation,
  $(a,b_0, D_u, D_v)$, are set such that the growth rates at the two Laplace--Beltrami eigenvalues,
  $\mu(\lambda_1)$ and $\mu(\lambda_2)$, both become zero.
  For positive $\nu$, the pattern can be approximated as
  \begin{align}
  {\bm U} = C_{1}(t) {\bm A}_{\lambda_1}W_{\lambda_1,k} + C_{2}(t) {\bm A}_{\lambda_2}W_{\lambda_2,k} + c.c.~,
        \label{eq:BrusselatorPatternApproximation}
  \end{align}
  where ${\bm A}_{\lambda_1}$ and ${\bm A}_{\lambda_2}$ are eigenvectors of the linear operator:
  \begin{align}
  	\mathcal{L}_0 =
	\begin{pmatrix}
		-\lambda_i D_u +b_0 -1 & a^2 \\
		-b_0 & -\lambda_i D_v -a^2
	\end{pmatrix}~.
  \end{align}
  Here, $\lambda_i~(i=1,2)$ are the eigenvalues of the Laplace--Beltrami operator for $W_{\lambda_i,k}$ and are determined from the surface shape. The eigenvalues $\lambda_i$ and parameters $(a,b_0, D_u, D_v)$ satisfy the relation explained in Supplemental Text Sec.~S3~\cite{Supplemental}.
  The eigenvectors ${\bm A}_{\lambda_i}$ and their conjugate vectors ${\bm B}_{\lambda_i}$ are expressed by Eqs.~(S52) and (S53),
  respectively.
  The coefficients of the amplitude equations are then calculated using the products of $\bm{A}_{\lambda_i}$ and $\bm{B}_{\lambda_i}$,
  and the integral with respect to $X_{\lambda_i,k}$.
  For more details,
  see Supplemental Text Sec.~S3~\cite{Supplemental}.

\subsection{Comparison between Brusselator model and amplitude equations \label{sec:AmpEqBrC}}
  We numerically solve the amplitude equations derived earlier,
  and we compare the solution with that of the Brusselator model.
  In particular, we explore the steady-state solution of the amplitude equations,
  which corresponds to either a static or propagating pattern,
  and confirm the claims given in Sec.~\ref{sec:AmpEqSte}.
  Dynamics other than the static and propagating patterns are discussed in the next section.

\subsubsection{Comparison procedure}
  The amplitude equations were derived using the following procedure.
  First, we fixed the surface and selected two pairs of Laplace--Beltrami eigenvalues,
  $\lambda_1$ and $\lambda_2$, that were adjacent in the spectral space,
  and then determined the corresponding eigenfunctions, $W_{1,\pm k}$ and $W_{2,\pm k}$.
  These were chosen to have the same wavenumber $k$.
  Next, we set the diffusion coefficients $D_u$ and $D_v$ and then determined the model parameters $a$ and $b_0$ to satisfy the determinant of $\mathcal{L}_0$ to zero (see Eqs.~(S58) and (S59)).
  Subsequently, we set the bifurcation parameter $\nu (=b-b_0)$,
  which was chosen to be sufficiently small and of the order of $10^{-2}$.
  The parameters were chosen so that the system would exhibit Turing instability,
  where the growth rates $\mu(\lambda_1)$ and $\mu(\lambda_2)$ were real and positive values.
  Finally,
  we determined the coefficients of the amplitude equations, $a_j$ and $b_j$ ($j=1,2,\ldots,7$), from Eqs.~(S43)--(S49), and performed numerical simulations.
  We also numerically solved the Brusselator model (the original system) using the same surface and parameters.

  We compared the solutions obtained from the amplitude equations and the Brusselater model in the phase space of $\eta$, $\xi$ and phase differences $\alpha$.
  For the Brusselator model,
  $(\eta, \xi, \alpha)$ was obtained by taking the inner product between the pattern $\bm{U}$ and eigenvectors $\bm{B}_{\lambda_1}W_{\lambda_1,\pm k}$ and $\bm{B}_{\lambda_2}W_{\lambda_2,\pm k}$, where $\bm{U}(t,x,\theta) =\bm{U}_{0}(t,x,\theta)+\mathcal{O}(|\nu|)$ and
  \begin{align}
    \bm{U}_{0}(t,x,\theta)
    &=\eta(t) e^{i\phi(t)}\bm{A}_{\lambda_1}W_{\lambda_1,k}(x,\theta) \nonumber\\
    &\quad+\xi(t) e^{i\psi(t)}\bm{A}_{\lambda_2}W_{\lambda_2,k}(x,\theta)+c.c.~,
  \end{align}
  where $\bm{A}_{\lambda_1}$ and $\bm{A}_{\lambda_2}$ are the same as the eigenvectors in the amplitude equations.

\subsubsection{Comparison of static and propagating patterns}
  The results of the numerical simulations for the amplitude equations and Brusselator model are summarized in Fig.~\ref{fig9},
  in which the blue and orange points represent the steady solutions $(\eta_0,\xi_0,\alpha_0)$ of the amplitude equations and Bursselator model, respectively.
  Patterns on axisymmetric surfaces and their projection onto the $x$-$\theta$ plane are also shown, where the left panels show the amplitude equations and
  the right panels show the Brusselator model.
  The coefficients of the amplitude equations are summarized in Supplemental Table~S1~\cite{Supplemental}.

  Figs.~\ref{fig9}(a) and (b) show static patterns of the steady solutions.
  For Fig.~\ref{fig9}(a),
  helical static patterns on the reflection-symmetric and $2\pi$-periodic surface are obtained through the direct simulation of the Brusselator model, while the constructed pattern from the amplitude equations exhibits a similar static pattern with phase difference $\alpha_0 =\pi/2$. For Fig.~\ref{fig9}(b), which employs the same parameter set as that in Fig.~\ref{fig6}(a),
  the orbit of the amplitude equations reach a solution similar to that of the corresponding Brusselator model, with phase difference $\alpha_0=0$.

  Figs.~\ref{fig9}(c) and (d) show propagating patterns with phase differences $\alpha_0 \neq n\pi/2$, respectively.
  For Fig.~\ref{fig9}(c),
  the helical propagating pattern is constructed from the amplitude equations for a surface without reflection-symmetry.
  $\alpha_0$ slightly deviates from $\pi/2$ and the pattern is propagating (see also Fig.~\ref{fig10}(c)).
  The patterns and dynamics obtained from the amplitude equations and Brusselator model are in close agreement.
  For Fig.~\ref{fig9}(d), the parameters in Fig.~\ref{fig5}(c) are used,
  where the phase difference is given by the non-trivial solution Eq.~(\ref{sol:reflec2}).
  We found that pattern dynamics constructed by the amplitude equations are similar to those of the original Brusselator model, indicating that the amplitude equations can capture the non-trivial propagating solution.

\subsubsection{Approximation accuracy against bifurcation parameter}
    The approximation accuracy against the bifurcation parameters of the helical propagating pattern was evaluated.
    Under the same conditions as those shown in Fig.~\ref{fig9}(c),
    the bifurcation parameter $\nu$ was changed from $0.01$ to $0.1$ in steps of $0.01$,
    resulting in the observation of similar helical patterns.
    %\magenta{\sout{through which similar helical patterns were observed.}}
    The patterns obtained from the amplitude equations closely approximated those of the Brusselator model, as shown in Figs.~\ref{fig10}(a), (b), and (c),
    in which the amplitudes $\eta$ and $\xi$ and the phase difference $\alpha$ of amplitude equations and the original Brusselator model are in close agreement.
    Figs.~\ref{fig10}(a) and (b) confirm that the leading order of the amplitudes is $\mathcal{O}(|\nu|^{3/2})$,
    which is assumed to derive the amplitude equations (Supplemental Text Sec.~S2~\cite{Supplemental}).
    The velocities obtained from the amplitude equations and the original Brusselator model
    are in close agreement (with difference $\mathcal{O}(\nu^3)$),
    and converge as the bifurcation parameter $\nu$ approaches zero (Fig.~\ref{fig10}(d)).
    Consequently,
    the numerical simulations confirm that the amplitude equations provide a reasonable approximation of the Brusselator model.

\section{Pattern dynamics other than propagation \label{sec:Dynamics}}
  In Sec.~\ref{sec:AmpEq} and Sec.~\ref{sec:AmpEqBr},
  we derived the amplitude equations for an axisymmetric surface and investigated the steady-state solutions corresponding to the static and propagating patterns.
  The generic form of the derived equations (Eqs.~(\ref{eq:amp_C1}) and (\ref{eq:amp_C2})) allows us to further explore the possible dynamics, other than propagation, of the RDE on general axisymmetric surfaces;
  the pattern dynamics are explored by numerically searching for non-stationary solutions of the amplitude equations.
  In the joint paper~\cite{joint}, we discuss this issue in detail;
  this section provides a summary of the results.
  As explained below,
  our results indicate that the Turing pattern can exhibit oscillatory and chaotic behaviors on curved surfaces.

\subsection{Limit cycle solution and pattern dynamics}
  We explored the possible pattern dynamics on curved surfaces by conducting numerical searches for amplitude equations (\ref{eq:amp_C1}) and (\ref{eq:amp_C2}).
  By changing the surface and model parameters of the Brusselator model,
  the corresponding amplitude equations were derived and numerically solved.
  For a given parameter set,
  the derivation of the amplitude equations is the same as that described in Sec.~\ref{sec:AmpEqBrC},
  where we used the surface $r(x) = d + \sum_{n=1}^{N} \left( s_{n}\sin(nx) + c_{n}\cos(nx) \right)$ with $ x\in[0,2\pi)$ and $N = 5$.
  For an efficient search,
  the range of the $x$-axis was narrowed.

  The surface shape and dispersion relations with parameters $(a, b_0, D_u, D_v) = (1.5, 2.97, 0.5, 2.2)$,
  $\nu = 0.02$, $d = 1, s_2 = 0.15, s_3 = 0.4$ and other surface parameters set to zero are shown in Fig.~\ref{fig11}(a) and (b), respectively.
  We found that the amplitude equations have a limit cycle solution, as represented by the blue line in Fig.~\ref{fig11}(c)
  (see Supplemental Table~S1 for the coefficients of the amplitude equations~\cite{Supplemental}).
  The corresponding pattern exhibits propagation accompanied by oscillations,
  as shown in Fig.~\ref{fig11}(d) left column.
  The bifurcation analysis suggests that the oscillatory pattern dynamics appear via Hopf bifurcation from the propagating pattern~\cite{joint}.
  We also validated that the original Brusselator model yields similar dynamics (Fig.\ref{fig11}(c) orange line and (d) right column).
  On a flat surface, the pattern remains static as an ordinary Turing pattern.
  This example demonstrates that the Turing pattern can exhibit more complex dynamics than propagation depending on the surface geometry.

\subsection{Chaotic dynamics in amplitude equations}
  We also searched for complex dynamics by changing the coefficients $a_j$ and $b_j$ ($j=2,3,\ldots,7$) as arbitrarily selectable parameters,
  with $a_1$ and $b_1$ chosen as real positive values to satisfy the condition of Turing instability.
  Note that, although this method does not identify the corresponding Brusselator model because of the difficulty in determining surface and model parameters from the coefficients,
  it allows for a general and efficient investigation of what pattern dynamics are possible.
  In addition to the limit cycle solution,
  we observed chaotic dynamics,
  as shown in Fig.~\ref{fig12}.
  The appearance of the chaotic dynamics can be understood in terms of period-doubling bifurcation against parameter $b_3$ (Fig.~\ref{fig12}(b)).
  The other coefficients of the amplitude equations are listed in Supplemental Table~S1~\cite{Supplemental}.
  The existence of such a chaotic solution in the amplitude equations suggests that rich and complex dynamics are possible depending on the geometry of the surface shape.
  However,
  the corresponding patterns have not yet been obtained for the Brusselator model on axisymmetric surfaces.

  As the search using the amplitude equations is limited to the region near the Turing bifurcation point on axisymmetric surfaces,
  we examined the Brusselator model under broader conditions.
  We found an example of chaotic pattern dynamics on a deformed spherical surface,
  as shown in Fig.~\ref{fig13}(a) and \cite{joint}.
  The surface is not axisymmetric and given by a radial function $R(\theta, \phi) = d + k (\cos (2\theta)-1) \cos(\theta) \cos(m\phi)$ in polar coordinate with $(d, k, m) = (6,0.206,2)$.
  The parameters for the Brusselator model were chosen to be the same as those used in Fig.~\ref{fig2},
  indicating that the pattern is static on a flat plane.
  We checked the initial condition sensitivity of the dynamics by comparing two time series of numerical simulations with slightly different initial conditions;
  the almost identical time series in the early stage eventually become distinguished in the late stage (Fig.~\ref{fig13}(b); see details \cite{joint}).
  These results show that chaotic pattern dynamics can arise due to the surface geometry.

\section{Discussion \label{sec:Discussion}}
  In our previous study,
  we reported that a Turing pattern that is static on a flat plane can propagate on curved surfaces~\cite{Nishide2022}.
  Using the relationship in Eq.~(\ref{eq:velocity_rel}),
  in which the pattern is assumed to move at a constant speed,
  we clarified that losses of symmetry in the surface and pattern,
  such as reflection asymmetry along the $x$-axis of an axisymmetric surface,
  are required for pattern propagation.

  In this study,
  we performed weakly nonlinear analysis as a complementary approach.
  This analysis allowed us to explore the generic dynamics of the Turing pattern on curved surfaces near the bifurcation point.
  The analysis was also advantageous in that the surface and pattern symmetries,
  which were treated separately in the previous study~\cite{Nishide2022},
  could be integrated using the eigenfunctions of the Laplace--Beltrami operator,
  which are determined by the surface shape.
  Because the mode decomposition of the helical and parallel patterns revealed that static and propagating patterns are described by the two paired modes,
  we derived the amplitude equations of two complex variables corresponding to the two paired modes.
  The obtained amplitude equations enabled the classification of pattern dynamics in the vicinity of the Turing bifurcation depending on the symmetry of the pattern and the surface shape.
  For a general axisymmetric surface,
  the steady-state solution of the amplitude equations,
  Eqs.~(\ref{eq:amp_C1}) and (\ref{eq:amp_C2}),
  enables pattern propagation.
  If a surface has reflection-symmetry or $2\pi$-periodicity along the $x$-axis,
  the amplitude equations depend on the combination of the modes.
  For helical and parallel patterns,
  the amplitude equations drop several terms and are reduced to Eqs.~(\ref{eq:C1_RS}) and (\ref{eq:C2_RS}),
  for which a propagation solution is no longer possible (see below for a nontrivial solution).
  On a flat plane,
  the amplitude equations are even simpler (Eqs.~(\ref{eq:AmpFlat1}) and (\ref{eq:AmpFlat2})),
  and the pattern caused by the Turing instability does not move.
  Overall,
  the interaction between the modes is more intense on general surfaces
  and leads to the onset of moving patterns.
  The propagation conditions given by the relationship between the surface and pattern symmetry are consistent with previous results.

\begin{figure}[h]
  \includegraphics[keepaspectratio,scale=1.0]{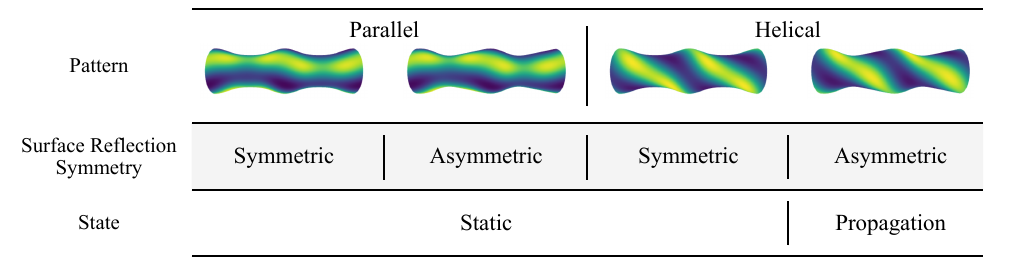}
  \caption{Relationships among pattern propagation, pattern profile, and surface symmetry reported in \cite{Nishide2022}.
  The helical pattern propagates on a reflection-asymmetric surface.
  The surface shapes are given by the radial function $r(x) =1.7+0.3\cos(x)+0.05\cos(2x)$ (for the reflection-symmetric surface)
  and $r(x) =1.7+0.3\cos(x)+0.05\sin(2x)$ (for the reflection-asymmetric surface with $-2\pi \leq x < 2\pi$).
  The color map represents $u(x,\theta)$,
  with a more intense yellow color indicating a higher chemical concentration.
  }
  \label{fig1}
  \end{figure}

\begin{figure}[h]
    \includegraphics[keepaspectratio,scale=1.0]{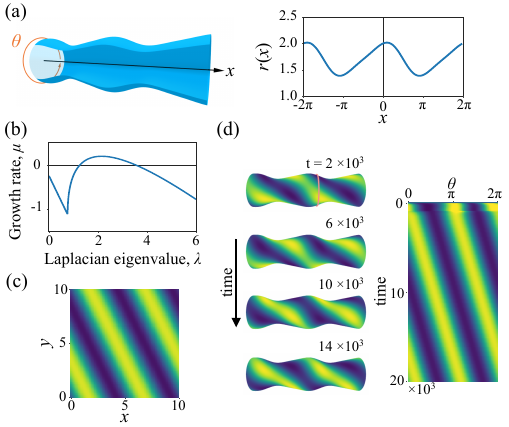}
    \caption{Propagating pattern on axisymmetric surface.
    (a) Axisymmetric surface; radial function $r(x)$ is shown in the right.
    (b) Dispersion relation of Brusselator model with parameter set $(a, b, D_u, D_v) = (2.0, 4.5, 0.5, 1.8)$.
    (c) Static pattern on a flat surface for Brusselator model.
    (d) Propagating pattern (left panel) and kymograph along the $\theta$ axis on the pale red line (right panel).
    The surface radius is $r(x)=d+k_1\cos(x)+k_2\sin(2x)$, where $(d,k_1,k_2) = (1.7,0.3,0.05)$.}
    \label{fig2}
    \end{figure}

\begin{figure}[h]
  \includegraphics[keepaspectratio,scale=1.0]{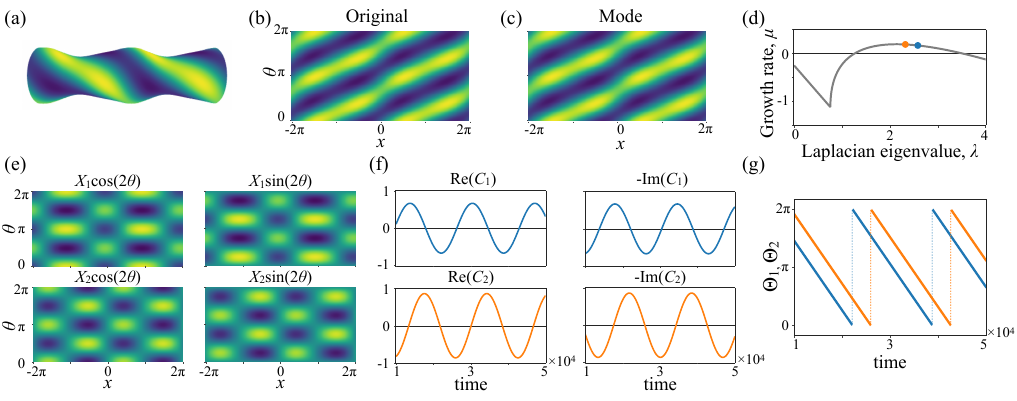}
  \caption{Mode decomposition of the helical propagating pattern.
   (a,b)~Helical propagating pattern and projected image on $x$-$\theta$ planes.
   The parameters are the same as those in Sec.~\ref{sec:propagating_pattern}.
   (c)~Approximated helical propagating pattern constructed using the two paired modes.
   (d)~Dispersion relation of the Brusselator model.
   The blue ($\lambda_1=2.58$) and orange ($\lambda_2=2.32$) points indicate the two paired modes of the approximation.
   (e)~Eigenfunctions of the two paired modes.
   $X_1(x)\cos(2\theta), X_1(x)\sin(2\theta), X_2(x)\cos(2\theta)$, and $X_2(x)\sin(2\theta)$.
   (f)~Time evolution of the amplitudes of the two paired modes. Real and imaginary parts of $C_1$ and $C_2$ are shown.
   (g)~Time evolution of phases $\Theta_1$ (blue) and $\Theta_2$ (orange) for the two paired modes.
   }
  \label{fig3}
  \end{figure}

\begin{figure}[t]
  \includegraphics[keepaspectratio,scale=1.0]{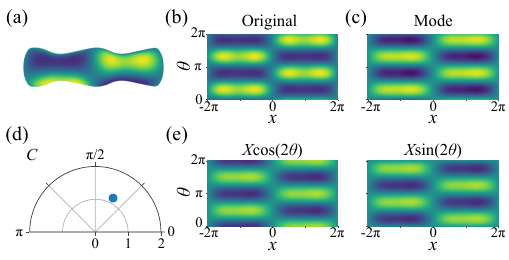}
  \caption{Pattern consisting of single paired mode.
  (a,b)~Static pattern and projected image on $x$-$\theta$ plane.
  We used parameter set $(a,b,D_u,D_v)=(1.7, 3.8, 0.5, 1.8)$
  and surface $r(x)=1.7+0.3\cos(x)+0.05\sin(2x)$.
  (c)~Approximated pattern constructed using a single paired mode.
  The Laplace--Beltrami eigenvalue of the mode is $\lambda=1.92$.
  (d) The amplitude $C=|C|e^{i\theta}$.
  (e) Laplace--Beltrami eigenfunctions $X_1\cos(2\theta)$ (left) and $X_2\sin(2\theta)$ (right).}
  \label{fig4}
  \end{figure}

\begin{figure}[t]
  \includegraphics[keepaspectratio,scale=1.0]{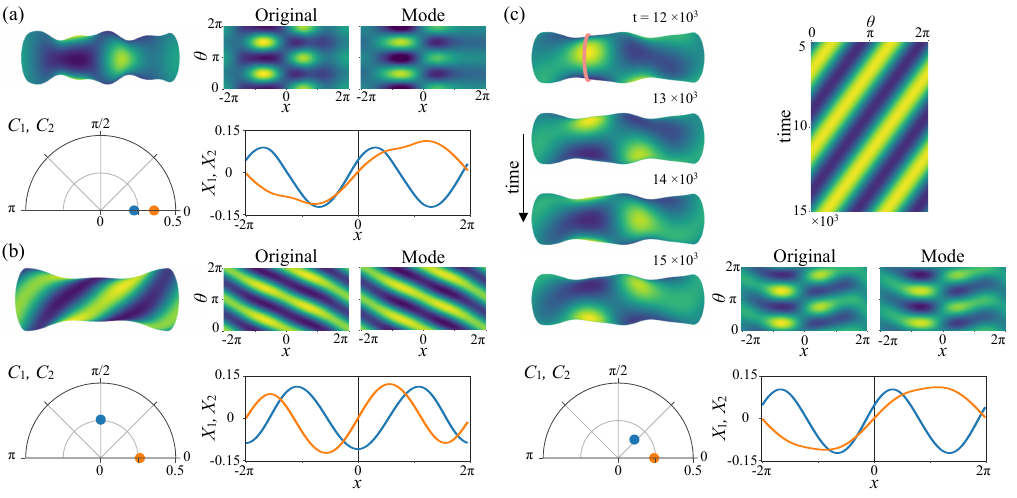}
  \caption{Patterns corresponding to the steady-state solutions of Eqs.~(\ref{eq:eta_reflection})--(\ref{eq:alpha_reflection}).
  (a--c) Patterns of the Brusselator model on axisymmetric surfaces,
  their projections onto the $x$-$\theta$ planes,
  and the approximated patterns constructed by eigenmodes.
  The amplitudes $C_1$ (blue) and $C_2$ (orange),
  and the corresponding eigenfunctions $X_1(x)$ and $X_2(x)$ are also shown.
   The phases of $C_2$ are set as $\Theta_2 = 0$.
  (c) Time evolution of the pattern and kymograph along the $\theta$ axis on the pale red line.
  Surface shapes $r(x)$ are described in the main text.
  (a)~Static pattern on $2\pi$-periodic surface. $(\lambda_1,\lambda_2)=(1.78,1.42)$, $(a,b,D_u,D_v)=(1.77,3.25,0.5,2.5)$.
  (b)~Static pattern on reflection-symmetric surface. $(\lambda_1,\lambda_2)=(1.84,1.86)$, $(a,b,D_u,D_v)=(2.07,3.73,0.5,2.5)$.
  (c)~Propagating pattern on $2\pi$-periodic surface. $(\lambda_1,\lambda_2)=(2.04,1.51)$, $(a,b,D_u,D_v)=(1.62,3.55,0.5,1.7)$.}
  \label{fig5}
  \end{figure}

\begin{figure}[t]
    \includegraphics[keepaspectratio,scale=1.0]{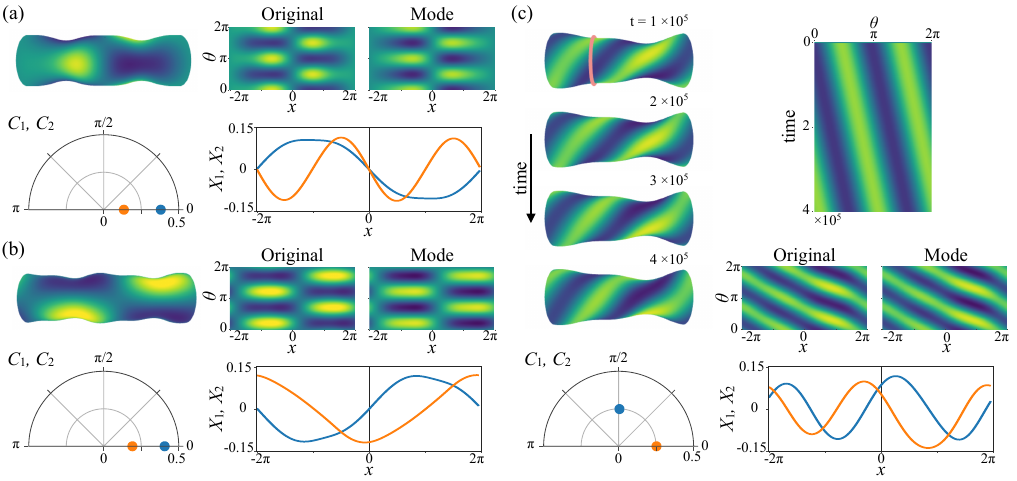}
    \caption{
    Patterns corresponding to the steady-state solutions of the amplitude equations in Eqs.~(\ref{eq:eta_general})--(\ref{eq:alpha_general}).
    The same configuration as Fig.~\ref{fig5}.
    Surface shapes $r(x)$ are described in the main text.
    (a)~Static pattern on reflection-symmetric surface. $(\lambda_1,\lambda_2)=(1.44,1.97)$, $(a,b,D_u,D_v)=(1.64, 3.43, 0.5, 1.9)$.
    (b)~Static pattern on $2\pi$-periodic surface. $(\lambda_1,\lambda_2)=(1.60,1.36)$, $(a,b,D_u,D_v)=(1.65, 3.05, 0.5, 2.5)$.
    (c)~Propagating pattern on general surface. $(\lambda_1,\lambda_2)=(1.83,1.91)$, $(a,b,D_u,D_v)=(2.09, 3.77, 0.5, 2.5)$.}
    \label{fig6}
    \end{figure}

\begin{figure}[t]
       \includegraphics[keepaspectratio,scale=1.0]{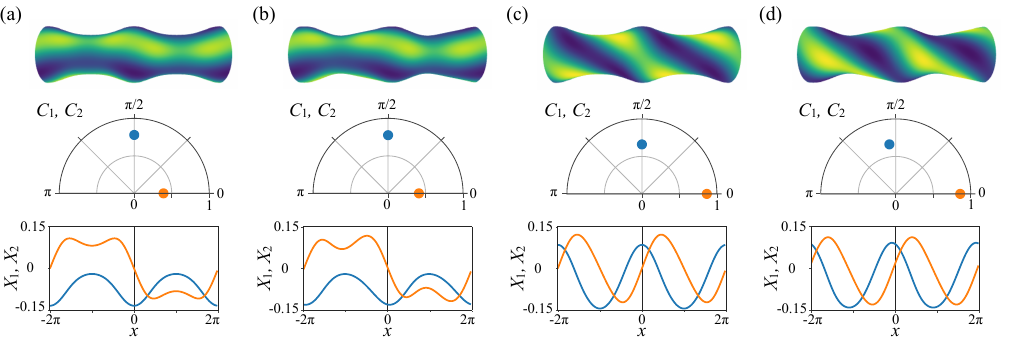}
       \caption{Eigenfunctions and phase differences of parallel and helical patterns.
       (a)~Parallel static pattern on $2\pi$-periodic and reflection-symmetric surface.
       (b)~Parallel static pattern on $2\pi$-periodic surface.
       (c)~Helical static pattern on $2\pi$-periodic and reflection-symmetric surface.
       (d)~Helical propagating pattern on $2\pi$-periodic surface.
       Surfaces are chosen as $r(x) = 1.7 + 0.3\cos(x) + 0.05\cos(2x)$ (a,c)
       and $r(x) = 1.7 + 0.3\cos(x) + 0.05\sin(2x)$ (b,d).
       Parameters are set as $(a, b, D_u, D_v)=(2.0, 4.5, 0.5, 1.8)$ (a,b)
       and $(2.8, 5, 0.4, 2.4)$ (c,d).
       See also Supplemental Figures S2--S4~\cite{Supplemental}.
       }
       \label{fig7}
    \end{figure}

\begin{figure}[t]
    \includegraphics[keepaspectratio,scale=1.0]{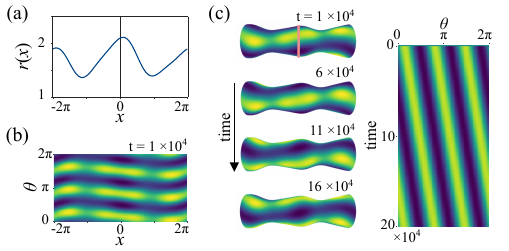}
    \caption{Propagation of a parallel pattern on a $2\pi$-aperiodic surface.
    (a)~Surface shape $r(x)=1.7+0.3\cos(x)+0.05\sin(2x)+0.1\cos(x/2)$.
    (b)~Pattern projected onto the $x$-$\theta$ plane.
    (c)~Time evolution of patterns on the surface and kymograph along the $\theta$-axis, represented by a pale red line.
    The parameters are set as $(a,b,D_u,D_v) = (2.8,5,0.4,2.4)$.
    }
    \label{fig8}

    \end{figure}

\begin{figure}[t]
    \includegraphics[keepaspectratio,scale=1.0]{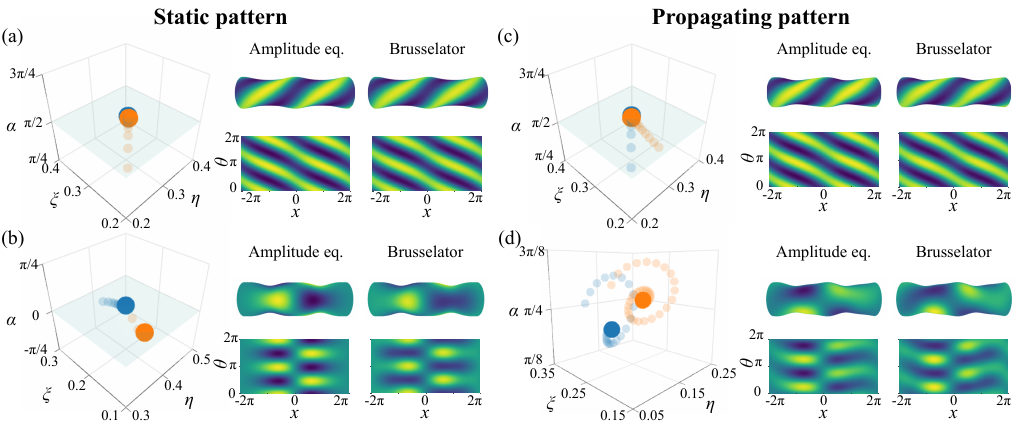}
    \caption{Comparison of steady solutions for the Brusselator model and amplitude equations.
    (a)--(d) Trajectories $(\eta,\xi,\alpha)$ for the amplitude equations (blue) and the Brusselator model (orange) are shown in the left panels,
    with steady-state solutions indicated by large points.
    Patterns on surfaces and their projected images onto $x$-$\theta$ planes are also shown for the amplitude equations and the original Brusselator model.
    Parameters are set as
    (a)~$r(x) = 1.7 + 0.18 \cos(x) + 0.03 \cos(2x)$, $(a, b_0, D_u, D_v) = (2.29, 4.86, 0.5, 1.8)$ and $\nu = 0.05$.
    (b)~$r(x) = 1.95 + 0.21 \cos(x) - 0.06 \cos(2x) + 0.015 \cos(x/2)$, $(a, b_0, D_u, D_v)=(1.64, 3.41, 0.5, 1.9)$ and $\nu = 0.02$.
    (c)~$r(x) = 1.7 + 0.18 \cos(x) + 0.03 \sin(2x)$, $(a, b_0, D_u, D_v) = (2.29, 4.86, 0.5, 1.8)$ and $\nu = 0.05$.
    (d)~$r(x) = 1.9 + 0.21 \cos(x) + 0.06 \sin(2x)$, $(a, b_0, D_u, D_v) = (1.62, 3.54, 0.5, 1.8)$ and $\nu = 0.01$.
    }
    \label{fig9}
    \end{figure}

\begin{figure}[t]
    \includegraphics[keepaspectratio,scale=1.0]{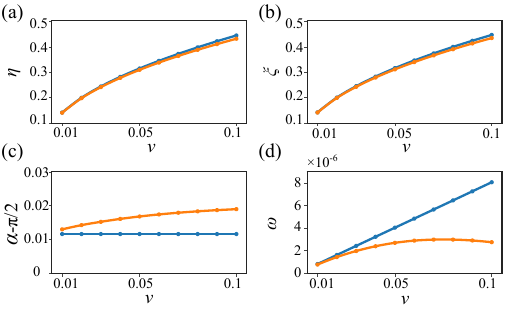}
    \caption{Approximation accuracy of amplitude equations for Brusselator model.
    (a) Amplitudes $\eta$ and (b) $\xi$, (c) phase difference $\alpha$, and
    (d) velocity $\omega$ are plotted against bifurcation parameter $\nu$.
    The blue and orange lines represent the amplitude equations and the original Brusselator model, respectively.}
    \label{fig10}
    \end{figure}

\begin{figure}[t]
    \includegraphics[keepaspectratio,scale=1.0]{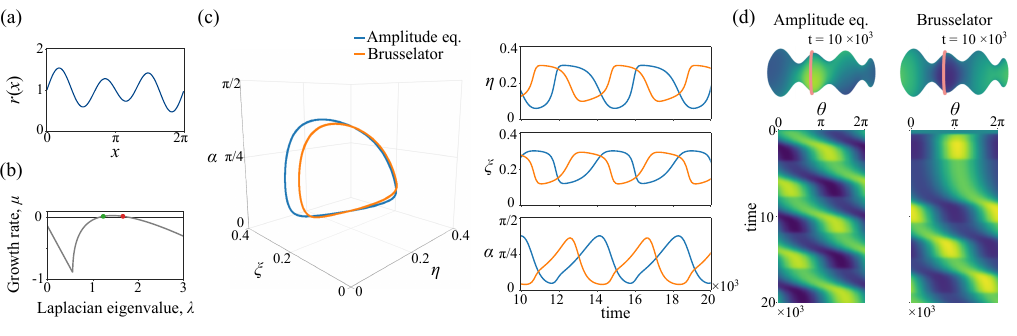}
    \caption{Limit cycle solution.
    (a)~Surface $r(x)=1+0.15\sin(2x)+0.4\sin(3x)$
    (b)~Dispersion relation and Laplace--Beltrami eigenvalues corresponding to the two paired modes (orange and green points).
    (c)~Trajectory of the limit cycle solution $(\eta,\xi,\alpha)$ (left) and time series of $\eta, \xi$ and $\alpha$ (right).
    The blue and orange lines represent the amplitude equations and the original Brusselator model, respectively.
    (d)~Snapshots of the limit cycle solutions and kymographs along the $\theta$-axis represented by pale red lines for the amplitude equations (left) and Brusselator model (right).
    See also Movie~1~\cite{Supplemental}.}
    \label{fig11}
    \end{figure}

\begin{figure}[t]
    \includegraphics[keepaspectratio,scale=1.0]{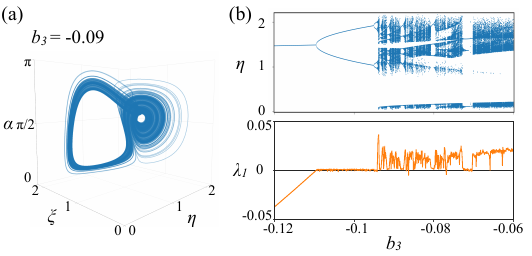}
     \caption{Chaotic dynamics in amplitude equations.
      (a)~Trajectory of chaotic solution at $b_3=-0.09$.
      (b)~Bifurcation diagram (top) and Lyapunov exponent (bottom) against $b_3$.
      For the bifurcation diagram,
      the value of $\eta$ on a Poincar\'e section is determined by
      $d\eta/dt=0$ was plotted for each value of $b_3$.}
     \label{fig12}
    \end{figure}

  The present study revealed several points beyond previous results.
  %\magenta{\sout{Compared with previous results,
  %the present study additionally revealed the following points.}}
  First,
  we found that the periodicity along the $x$-axis is involved in determining the pattern dynamics.
  The parallel pattern does not move on the $2\pi$-periodic surface (Fig.~\ref{fig1}),
  whereas the loss of periodicity causes pattern propagation (Fig.~\ref{fig8}).
  Second,
  the pattern can propagate even on a surface with reflection-symmetry and $2\pi$-periodicity,
  corresponding to the nontrivial steady-state solution of the amplitude equations.
  For such a solution, the pattern profile is not symmetric.
  This type of solution is consistent with the propagating pattern on a sphere determined numerically in the previous study~\cite{Nishide2022}.
  Third,
  a weakly nonlinear analysis revealed that patterns on axisymmetric surfaces exhibited more complex dynamics than propagation,
  including oscillating and chaotic chemical waves.
  Some of these dynamics were confirmed through numerical simulations of the RDE for the Brusselator model~(Fig.~\ref{fig11}).
  In particular,
  motivated by the chaotic solutions in the amplitude equations,
  we found a chaotic Turing pattern by direct numerical simulation of Brusselator model on a deformed spherical surface (see the joint
  paper~\cite{joint}).
  These findings provide a new perspective on the intricate interplay between surface characteristics and pattern dynamics.

\begin{figure}[t]
    \includegraphics[keepaspectratio,scale=1.0]{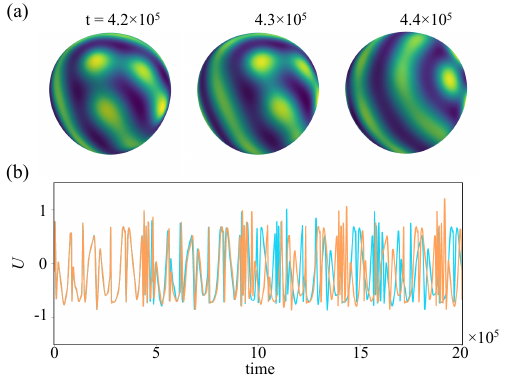}
     \caption{
     Chaotic dynamics of Turing pattern on a deformed sphere.
      (a)~Time evolution of Turing pattern.
      (b)~Time series of chemical concentrations $U = u-a$ at a point of the surface.
      The orange and light blue lines represent two different time series
      with slightly different initial conditions.
      See the joint
      paper~\cite{joint} for details.
      }
      \label{fig13}
    \end{figure}

  The present study should be extended to pattern dynamics on general surfaces
  without axisymmetry, for which weakly nonlinear analysis is applicable.
  By performing numerical simulations of the RDE on several surfaces,
  we observed chaotic patterns (Fig.~\ref{fig13}) and moving patterns with varying speeds (see Movie~2~\cite{Supplemental}).
  Here,
  we briefly consider these patterns in terms of the amplitude equations:
  without axisymmetry (i.e., no translational invariance along the $\theta$-axis),
  amplitude equations can have quadratic terms representing additional interactions among modes,
  and more complex dynamics could arise.
  A detailed analysis is required in the future.

  Pattern dynamics driven by surface curvature are applicable to natural and engineering systems.
  By controlling the parameters of the surfaces and reactions, the pattern can be switched between oscillating, static, and propagating states.
  In biological systems,
  such dynamics can be used for information transduction,
  guidance of cellular migration,
  and positioning of molecular localization.
  Analyses using amplitude equations are useful for predicting pattern dynamics and can also provide a basis for the engineering of patterns on curved surfaces.
  In the future,
  analysis of the stability and bifurcation of the amplitude equations will further improve our understanding of pattern dynamics on general curved surfaces,
  and will also be useful for dynamics in network systems~\cite{Nakao2010, vanderKolk2023emergence},
  and on deformable surfaces~\cite{toole2013turing, Miller2018, Tamemoto2020}.

  \vspace{8mm}
  \noindent{\bf Acknowledgements}\\
  This study was supported by
  JST SPRING JPMJSP2108 (to R.N.),
  JSPS KAKENHI JP23KJ0641 (to R.N.),
  JPJSJPR 20191501 (to S.I.),
  and JST CREST JPMJCR1923,  Japan (to S.I.).

  \vspace{5mm}
  \noindent{\bf Author contributions}\\
  R.N. and S.I. both proposed the research direction, contributed to the theoretical analysis,
  and wrote the manuscript. R.N. performed all numerical simulations.

\end{document}